\documentclass[preprint,10pt,onecolumn]{elsarticle}
\usepackage[dvips]{epsfig}
\usepackage[dvips]{graphicx}
\usepackage{graphicx}
\usepackage{latexsym}
\usepackage{amssymb}
\journal{Journal of Systems Architecture}
\begin{document}
\begin{frontmatter}

\title{How to Build Vehicular Ad-hoc Networks on Smartphones}
\author{ Pino Caballero-Gil, C\'andido Caballero-Gil, Jezabel Molina-Gil} 
\address{Department of Statistics, Operations Research and Computing,  University of La Laguna, Spain. E-mail addresses: \{pcaballe, ccabgil, jmmolina\}@ull.es}

\begin{abstract}
Vehicular ad-hoc networks have been defined in the literature as communications networks that allow  disseminating information among vehicles to help to reduce traffic accidents and congestions. The practical deployment of such networks has been delayed mainly due to  economic and technical  issues. This paper  describes a new  software application to detect traffic incidents and exchange information about them, using only smartphones, without any central authority or additional equipment. Both road safety and  communication security   have been taken into account in the application design. On the one hand, the interface  has been designed  to avoid distractions while driving because it operates automatically and independently of the driver, through voice prompts. On the other hand, communication security, which is essential in critical wireless networks, is provided  through the protection of attributes such as authenticity, privacy, integrity and non-repudiation. All this is achieved without increasing the price of vehicles and without requiring the integration of new devices neither in vehicles nor on roads. The only prerequisite is to have a smartphone equipped with Wi-Fi connectivity and GPS location in each vehicle. The proposed application has been successfully  validated both in large-scale NS-2 simulations and in small-scale real tests to detect traffic congestions and empty parking spaces. 
\end{abstract}

\end{frontmatter}

\section{Introduction}
\label{intro} Vehicular Ad-hoc NETworks (VANETs) are a form of
Mobile Ad-hoc NETworks (MANETs) that provide the infrastructure
for developing new systems to improve safety and comfort of driving. VANETs are  generally considered formed by mobile nodes
corresponding to On Board Units (OBUs) in vehicles, and stationary
nodes called Road Side Units (RSUs) in the infrastructure of the
road. Many efforts are being made to define new standards for
 services and interfaces of VANETs. In \cite{WAVE} the
architecture called Wireless Access in Vehicular Environment
(WAVE) based on the IEEE 802.11p and the IEEE 1609 standards was
published. However, to the best of our knowledge, no large-scale real deployment of such a standard  has been performed yet.

Currently, there are several research projects  on VANETs
\cite{VAN11}, and many enterprises  are developing different
services, which  VANETs could solve if they were already deployed. Among
the  potential applications of VANETs, the most remarkable one is the dissemination
of traffic congestion information  and collision warnings, but that is not the only one.
Many others exist such as parking
availability notification, vehicle tracking, weather information, advertising,
Internet provision, etc.

Dissemination of traffic congestion information  is regarded as one of the
most important applications of VANETs because the number of traffic congestions  increases as the number of vehicles
 grows.  Apart from negative effects on economy, traffic congestions produce high levels of
stress in people and are the major cause of air pollution. It has
been shown that people caught in traffic are three times more
likely to have a heart attack than those who are not stuck in a
congestion. Communication among vehicles could help to prevent this
problem by reducing traffic congestions, what would also avoid enormous
wastes of time, money, and resources.

This paper describes VAiPho, a secure communication system for spontaneous and self-organized vehicular networks based on smartphones with GPS and Wi-Fi connectivity, which does not require any infrastructure in vehicles or on roads because its operating mode is completely distributed and decentralized. 
In particular, communications in our proposal are  based only on  Wi-Fi in order to reduce the cost to zero, because  existing smartphones  offer that type of connectivity, and their use has no cost. Although other communication techniques such as 3G, WiMAX, 4G, etc. can provide higher transmission speed, longer transmission distance and larger network throughput, one of their main problems is that they are not available everywhere. Besides, existing solutions based on those techniques imply that users may have to change their phones,  pay for use, and/or lose their  privacy. Furthermore, our proposal is based on 802.11b because it is thought for its use with reduced speeds in urban environments where traffic congestions and lack of empty parking spaces are problems that need urgent solutions. 

The main goal of VAiPho is to increase safety and comfort of driving through the exchange of warning messages about traffic congestions. It also allows taking advantage of additional services such as empty parking space detection, parking reminder and geo-located advertising. One of its main features is that it is a secure system because it protects user privacy and data integrity. 

This paper is organized as follows. Section~\ref{sec:Related} covers some related research on security and applications of VANETs. The technology required by the system specifications is presented in Section~\ref{sec:Requirements}. Section~\ref{sec:Structure} contains a detailed description of VAiPho structure, including explanations of its main applications. Section~\ref{sec:Website} describes the VAiPho items that are required for the management of network trust. Security issues related to information content and user anonymity are analyzed  in Section~\ref{sec:Security}, while some implementation results are provided in Section ~\ref{sec:Result}. Finally, conclusions and future work are presented in Section~\ref{sec:Conclusion}.

\section{Related Research}
\label{sec:Related}
A recent survey on research in vehicular ad-hoc networks is provided in \cite{Zeadally}, where the authors present a review of wireless access standards,  trials and simulators of VANETs.

When designing a tool to create a self-organized vehicular network with the goal of increasing road safety, the first prerequisite to be considered is the accuracy and reliability of transmitted information. Thus, security is the most important topic to be taken into account when a communication system is designed for VANETs \cite{Ray07}. In the bibliography we can find several proposed schemes for self-organization in VANETs \cite{Wis07}, MANETs \cite{Cap03}, and sensor networks \cite{Sohrabi}, which try to solve all or part of the security problems in those types of networks. However, a different approach is presented in this paper, where a  self-organized, and at the same  time, practical and  secure  way to deploy VANETs  is proposed. 

An especially important security aspect of the system is user privacy. Our proposal uses a random pseudonym generator to guarantee with  high probability that it is not possible to track a vehicle, and that coincidences between two generated pseudonyms are very unlikely.  
The paper \cite{But07} proposes a specific pseudonym-based scheme  to solve the privacy problem  caused when GPS coordinates and speeds of  vehicles are sent in the beacons. In our proposal, none of those data are sent in beacons. On the other hand,  to cope with the issue of changing pseudonyms in improper times or locations,  either the mix model \cite{Gerlach} or social spots \cite{Lu12} might be combined with our proposal.

With respect to the general objective of  discovering and disseminating traffic congestion information,  the work \cite{Dor07} has the same goal of this work, but it does not address the important aspect of security of connections. Also many existing centralized GPS software applications offer traffic services based on information provided by  local road authorities, police departments and systems that track traffic flow.  However, neither of
them  are  real-time data as they do not reflect the
events that have just produced, nor respect users' privacy.
 For instance, Google Traffic
\cite{GOOGLETRAFFIC}, TomTom \cite{TOMTOM}, Sygic \cite{Sygic} and
Waze \cite{WAZE} are well-known solutions to detect traffic jams. The main
difference with our proposal is that all of them need  mobile data connection. Another disadvantage is that
users completely lose their privacy because they have to provide
information about their locations to the companies and other bodies that support the service.

 Regarding the search for empty parking spaces  a few solutions exist but, to the best of our knowledge, none of them is based simply on  a mobile application. The paper \cite{Mat10} presents a proposal where through a device installed in the passenger door, the empty parking space is found and reported to a centralized server through  3G or GPRS.
 In \cite{Pan07} the authors propose a solution where users can find empty parking spaces managed by RSUs. The paper
 \cite{Caliskan}  proposes a dissemination algorithm for spatio-temporal traffic information such as parking space availability, but its goal  is not on how the information is obtained, but on how it is transmitted through the network.

The solution to find the parked car is the easiest to implement and consequently several mobile applications can be found for different mobile platforms \cite{CFin}  \cite{CarFin} \cite{CarSpo}   \cite{LGPar} \cite{LocGPS} \cite{GPark} \cite{Where}. Anyway, such a use of the application is simply a value-added feature for the proposal, and not its main goal.

This paper takes into account that the introduction of a complete model of WAVE-based VANETs is extremely expensive both for users, who would have to buy and install new devices for their vehicles, and for the state, which would have to deploy a huge infrastructure to support VANET services. Therefore, this work proposes a self-organized VANET without any infrastructure, which  serves as introduction to a more complex  VANET, all this with good levels of reliability and security.
Our main goal is to define a simple and scalable model for VANETs where users can cooperate through their mobile devices and obtain updated information of interest about their traffic area in order to choose the best  route to their destinations. Our proposal takes into account that the integration of VANETs will be gradual, so that at the beginning there will not be any RSU, and the VANET will start with only a few vehicles. This growth will be faster or slower depending on the popularity, acceptance and ease of use of VANETs.
Thus,  this paper focuses on the first phase of the deployment. As soon as WAVE-based VANET infrastructures are fully deployed, the proposed solution should be checked to avoid any unnecessary communication.

\section{Design Requirements}
\label{sec:Requirements}
Mobile application development has gone multiplatform, so the system here proposed, called VAiPho, has been tested for Android, Windows Mobile and Symbian, and is being developed for iOS and Windows Phone.

The minimum system specifications (see Fig. \ref{fig:connections})  for the optimal use of VAiPho are the following:

\begin{figure*}[]
\centering
 \includegraphics[width=0.70\textwidth]{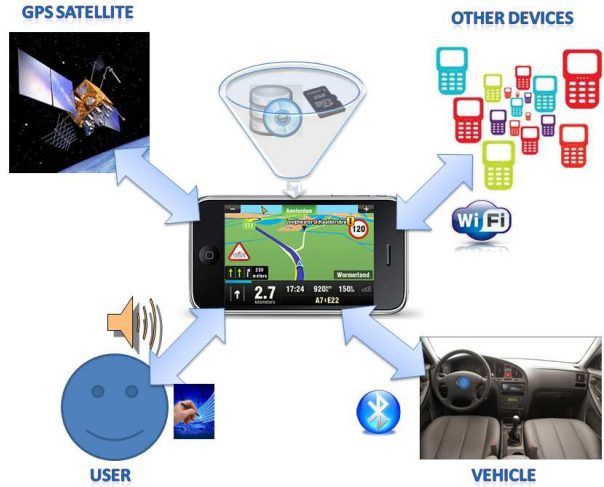}
\caption{Minimum System Specifications}
\label{fig:connections}       
\end{figure*}

\begin{itemize}
\item Bluetooth: To  connect the device with the vehicle, providing automatic activation of VAiPho without requiring the user's attention.
\item Wi-Fi IEEE 802.11 b/g: To allow free exchange of information about possible events between  wireless devices.
\item GPS antenna: To get the GPS coordinates where the events happen and the speed and direction of the vehicle.
\item Storage space: To provide enough capacity for storing programs and data.
\item Database: To manage the storage of user data and information about different events such as possible empty parking spaces or reminders of parked vehicles.
\end {itemize}

VAiPho takes advantage of the powerful application platform capabilities of today's mobile devices. It has no cost at all neither for users nor for governments because it does not require any RSU, and the OBU's role   is played by the smartphone inside the vehicle. Furthermore, since people are used to  smartphones, the interface is not strange for them, what avoids usual difficulties of learning to deal with new devices.

VAiPho  communications are performed using the wireless IEEE 802.11b/g  standard. It is well-known that in general  this standard is not the best suited for road safety applications because communications, which are in the 2.4 GHz range, may interfere with other devices such as Bluetooth, Zigbee or other WLAN networks in the same channel. Furthermore, other problems exist, such as delays and  limitations in retransmissions, non-directional  antennas, battery consumption and difficulties caused by different speeds of vehicles. All these problems occur because such a standard was not defined for VANETs. Instead, the IEEE 802.11p standard of wireless communications  \cite{Jia08} was specifically designed for the deployment of VANETs and therefore it is more appropriate for this type of communications.
However, currently there are no devices capable of broadcasting in the frequency range that IEEE 802.11p uses. In contrast,  most  smartphones have the capability to communicate under the IEEE 802.11b/g standard. Thus, VAiPho can be seen as a practical tool for deploying VANETs with today's mobile phones.
Furthermore, VAiPho has been designed in modules, so the communication module might be easily adapted to the IEEE802.11p standard  in the future when this standard is widespread.

Many communication tests  have been made using the IEEE802.11b/g standard with VAiPho in mobile phones inside cars. The first conclusion of these tests is that  vehicles  traveling at high speeds in opposite directions  do not have enough time to communicate.  However, vehicles in the same direction or inside cities, where the speed is lower, have time enough to establish communications and to exchange the required data.

Thus, the system described here can be seen as a first approach to the real deployment of some interesting VANET applications with currents devices, what could be analyzed as a proof of concept of the future WAVE-based VANETs.

\section{VAiPho Architecture}
\label{sec:Structure}
\begin{figure*}
\centering
 \includegraphics[width=0.50\textwidth]{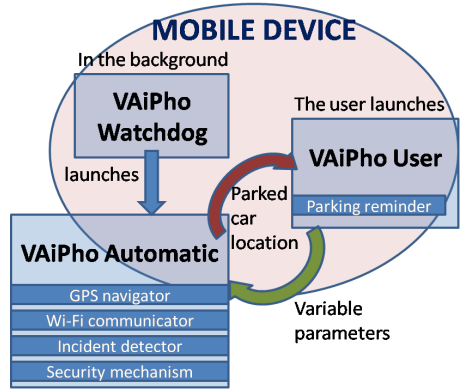}
\caption{VAiPho Architecture}
\label{fig: structure}       
\end{figure*}

VAiPho is composed of three main modules, as shown in Fig. \ref{fig: structure}:
\begin{itemize}
    \item \textit{VAiPho Watchdog}, which runs in the background, is a light program that automatically connects the mobile phone to the car hands-free system.
    \item \textit{VAiPho Automatic},  which starts the automatic application that is responsible for detecting and forwarding the events that happen on the road.
    \item \textit{VAiPho User}, which is specially designed to provide interesting functionalities to the users, allowing them to interact with VAiPho.
\end{itemize}

This section describes the internal structure of these three modules.

\subsection{VAiPho Watchdog}
\label{sec:Background}
One of the most important issues when designing applications for road safety is that users keep their eyes on the road, because distracted driving is a major cause of death on the road. For this reason we propose \textit{VAiPho Watchdog}, which is a mobile application that  is continuously listening to the registry so that when the mobile phone connects to  the car through Bluetooth, 
it automatically launches the  application. In particular, \textit{VAiPho Watchdog} starts  automatically when the user switches on the mobile phone,  and then it runs in the background.  It hardly consumes any resources because its unique function is to listen to a registry and to launch \textit{VAiPho Automatic} if the  device connects to the car through Bluetooth.

A potential drawback of VAiPho is that it uses multiple communication interfaces at the same time, which could imply high battery consumption. Anyway, the most battery consuming utility of mobile phones is data connectivity, which is not used in VAiPho. Since we are aware that the priority of mobile phones is to make phone calls, and VAiPho battery consumption could cause discomfort to users,  to solve this problem, \textit{VAiPho Watchdog} checks battery level before launching \textit{VAiPho Automatic}. In order to carry out this process, it distinguishes among five possible states of the phone battery: very high, high, medium, low and very low. If the battery level reaches a certain threshold, it does not launch \textit{VAiPho Automatic}.  Users  can choose their preferred threshold from those five possibilities. The default value is set to low.

\subsection{VAiPho Automatic}
\label{sec:Automatic}
The main application  of VAiPho is \textit{VAiPho Automatic}, which requires  connectivity with other phones via Wi-Fi. Besides, it uses GPS information, like geographic coordinates, vehicle speed or speed limits in order to detect possible traffic congestions or empty parking spaces. All this information is processed, stored and sent to other vehicles by \textit{VAiPho Automatic}. For that purpose this module implements the following functions:
\begin{itemize}
\item Starting the wireless interface
\item Setting up, or connecting to the ad-hoc network called \textit{VAiPho}
\item Loading data from the database and  from the password file
\item Loading and starting the client beaconing and the server
\item Starting the GPS navigator
\item Starting the incident detector
\end{itemize}

When  \textit{VAiPho Automatic} is launched, it always follows the same process. First, the mobile wireless interface is started and a new ad-hoc network called \textit{vaipho} is created, or if it already exists, the device connects to it. Then, it creates and fills the database with user data such as public/private key pairs, secret keys, pseudonyms, etc. These  security issues are detailed in Section \ref{sec:Security}. Then,  client and server processes are started to receive and send beacons. At this point, the system is ready to exchange information about events with other devices in the range of transmission. This entire process is automatic and transparent to the user, and just a voice message indicates to the driver that VAiPho has begun.

After setting the communication system, the GPS navigator is started and an incident detector  is launched. The main aim of this system is to detect any anomalous situation in order  to generate the corresponding event warning  to alert other users about that situation. This process uses GPS software to obtain  data such as speed and location. Specifically for the real device implementation of VAiPho we  used the Sygic GPS navigation Software Development Kit (SDK) \cite{Sygic} that allows developers to add navigation features to  mobile applications. The main goal of \textit{VAiPho Automatic}  is to  detect traffic congestions  automatically, so  it uses the function of the SDK that  allows retrieving   location and speed information as well as  speed limit of the  road. With this information, the event detector of VAiPho finds out whether the vehicle is travelling at an abnormally low speed and in such a case, it concludes that the vehicle can be in traffic congestion on the road. Once detected the incident, the process generates an event warning with the road name, direction of movement and location in which the incident is located. This event is stored in the database and relayed to other vehicles. Thanks to these event detections, and through cooperation among devices, it is possible to know more about road conditions.

When a vehicle receives this information, two different cases can occur. The first  case consists in that the vehicle is travelling on the same road and also detects the same event. In this case, we use a data aggregation scheme in order to avoid network overload. This  issue is explained in Section \ref{sec:Security}. The second case corresponds to a vehicle that does not detect the event because either it is not circulating on the same road where the congestion has been detected, or it is on the same road but at a high distance from the incident. In this case two possibilities can be distinguished:
\begin{itemize}
\item The vehicle  does not have that road as part of its route: In this case, the vehicle acts as a simple packet transmitter and relays the packet to all vehicles in its range of transmission.
\item The vehicle has that road as part of its route: This case  takes more processing time because the vehicle has to first determine whether the busy road is in its route, and if so, it has to compute whether it is better to choose an alternative route or to keep the same one. If the system determines that it is better to change the route, the new route is computed.
\end{itemize}

In order to implement all these processes, many security issues have had to be taken into account because it is easy to try to generate false events, or to modify the contents of true packets, or even to deny relaying packets, trying to attack the network operation. Therefore, communications among vehicles and information about detected events relayed in the network should provide evidence of being truthful. For this purpose, a security module that will be presented in Section \ref{sec:Security} has been created. There, we explain several possible attacks and how VAiPho resists all of them.

Another important task of  \textit{VAiPho Automatic} is  the detection of  empty parking spaces. In this tool, the procedure is simple. When a driver turns on the car, his/her mobile device gets synchronized to the GPS signal and obtains the geographic coordinates where the vehicle is parked before it leaves the space. Then, these geographic coordinates are broadcasted by \textit{VAiPho Automatic} as an empty parking space. Received events of parking spaces have a short expiration time that is configurable by the user. The default value is set to 1 minute in the implementation. Frauds and errors cannot be controlled for parking events because there are several situations where a vehicle can leave a certain location that is not an actual empty  parking space, such as a private outdoor parking space,
 a  double parking, or
 a prohibited parking.
For this reason, when the tool announces  a space, it really indicates that it is a potential empty parking space, but there is no guarantee that the space continues to be available when the receiver reaches it. In order to use this tool, it is necessary to launch the search for an empty parking space. A voice message is then launched when an empty parking space is detected and a parking signal is shown on the map.

On the other hand, when the user turns off the car, the geographic coordinates where the vehicle is parked are stored in the database in order to help the user to find his/her parked vehicle. Often,  it is difficult for users to find the places where they parked their cars, especially in large and unknown cities. In these cases, VAiPho uses the parked vehicle's coordinates stored in the database to draw a walking  route to the car on the map.

\begin{figure}
\begin{center}
\begin{tabular}{c}
\includegraphics[width=0.2\textwidth]{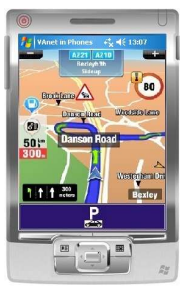}

\end{tabular}
\label{VAiPho Interfaces}
\caption{  Driver Interface of \textit{VAiPho Automatic}}
\end{center}
\end{figure}

The interface that is displayed when the user is driving (see Fig. 3) is very similar to that used by conventional GPS navigation applications.  In order to avoid driver distraction, this interface not only uses  icons on the maps but also voice messages. Therefore, when it detects traffic congestion, an icon on the screen is shown and a voice message indicating congestion on the route is heard. The same method is used to report a possible empty parking space.

\subsection{VAiPho User}
\label{sec:UserApplication}

\begin{figure}
\begin{center}
\begin{tabular}{c c}
\includegraphics[width=0.2\textwidth]{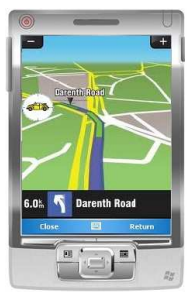}
\end{tabular}
\caption{ \label{VAiPho Interfaces2}First Pedestrian Interface of \textit{VAiPho User}}
\end{center}
\end{figure}

\begin{figure}
\begin{center}
\begin{tabular}{c c}
\includegraphics[width=0.2\textwidth]{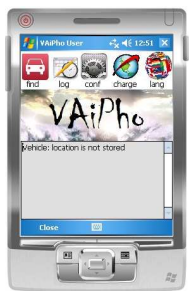} 
\end{tabular}
\caption{ \label{VAiPho Interfaces3}Second Pedestrian Interface of \textit{VAiPho User}}
\end{center}
\end{figure}

A good user interface design can be the difference between product acceptance and rejection in the market. 
We have designed two simple interfaces for \textit{VAiPho User } module, which are shown in Fig. \ref{VAiPho Interfaces2} and Fig. \ref{VAiPho Interfaces3} when the user is not driving. In this module,  the user can configure variable parameters like parking expiration, battery level threshold, and application language or the trust data  generated through the VAiPho website that will be described in the next section. Additionally, the user can check the events that are updated and stored in their mobile phones. In particular, \textit{VAiPho User } module provides a tool that allows  locating where the vehicle is parked if  \textit{VAiPho Automatic} was active when the car was parked and if the parking location  had GPS coverage. For this purpose \textit{VAiPho User} module provides a parking reminder button marked with word \textit{find}, which the users have to  click on to locate their cars. This module  starts the GPS navigator on walking mode to show a walking route on the map between the  current location and the parked car. Fig. 4 and 5 show respectively the  cases when a message indicates that  information on the vehicle location is and is not  available.

\section{VAiPho Website}
\label{sec:Website}

VAiPho website \cite{VaiWeb}  is useful to advertise the application  and  to allow  downloading the tool. However, its importance comes mainly from its use  for defining trust relationships between users in order to define the trust graph according to which key certificates are distributed. The main page has a menu from which  points of interest,  such as \textit{Downloads}, are highlighted. From there, users can download  VAiPho for different platforms and update it to any new features that  have been developed. In the \textit{Users} menu, people interested in VAiPho must register, what allows   generating a file of signatures and certificates that is required to use the application.  In addition,  a support service is provided in the \textit{Main} section, where users who install the application can raise  questions about its operation. Finally, in the \textit{Clients} section, companies or users who are interested in geolocation advertising through this tool may contact the administrators. The importance of this feature is because this is the source of commercial profit.  The advertising data include name of the company, message, X coordinate, Y coordinate, area of interest, expiration date and the commercial logo. The message is limited to a few characters in order to allow its easy reading.
This information is processed and checked, and after payment, the user  receives a packet  containing the advertising information and corresponding certificate. Then, the user downloads \textit{VAiPho Advertising Application} and puts it in a predetermined path in the device. Afterwards, the device starts broadcasting  the advertising information, and nearby VAiPho users receives it if it is defined in their advertising filters.

\begin{figure*}
\centering
  \includegraphics[width=0.77\textwidth]{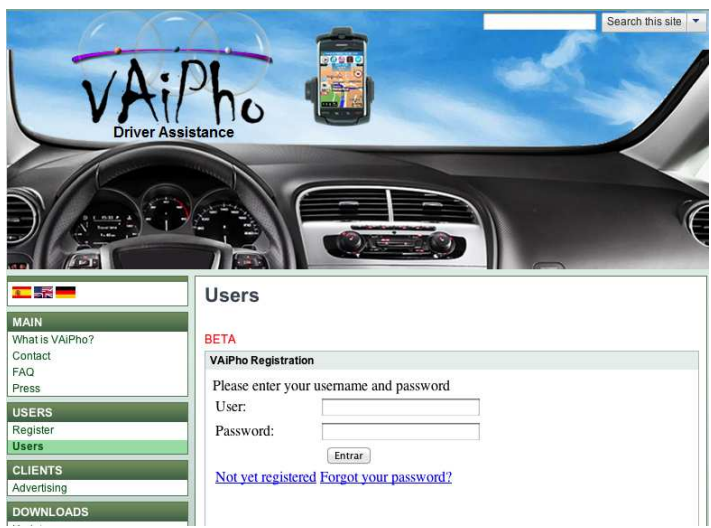}
\caption{VAiPho Website}
\label{fig:Website}
\end{figure*}

VAiPho is a self-organized tool based on the trust between users represented with key certificates signed the users who trust the corresponding key and owner. The more signatures a key certificate has received, the greater the trust from other nodes that can be used during its life in the network. Therefore, we can find a similarity between VAiPho network and social networking sites like Facebook, Twitter, etc. In this way, VAiPho operates according to what is called viral marketing, that is to say, it uses  techniques based on pre-existing social networks.

With this in mind, when a user logs in the web  (see Fig. 6) and registers  to get his/her  pair of public and private keys in order to start using VAiPho, the system asks him/her for information about his/her contacts in some social network. Currently, the tool can import contacts from a mail account. In this way, VAiPho checks this list looking for any  users also registered in the system. The results are displayed to the new user so that he/she can confirm which of these users are chosen to sign his/her certificate.  During the signing process, \textit{VAiPho Website} exchanges the generated certificates so that each user holds the signature generated by him/herself and the signatures produced by his/her friends.  Each time  a new mutual authentication is made, the user may download an updated file containing the certificates generated by him/herself and his/her friends, which is needed to generate the certificate graph of the trust network in order to allow secure communications along the network. This process and the place where the file has to be stored are detailed  in the \textit{Users} menu of the VAiPho Website.

From the \textit{Download} menu,  users must download the application according to their OS platform, install the tool on their device and put the generated files on the indicated path. When the device starts VAiPho, the program updates the database with the information contained in the aforementioned files.

\section{Security Analysis}
\label{sec:Security}

Security of communications in VANETs represents an important challenge because in these networks warning messages affect driver decisions  about  reducing speed and/or choosing alternative routes based on the received information.  Thus, a security  scheme is necessary to determine whether the road traffic information available to the driver is trustworthy or not. Besides, the quality of communications in VANETs is degraded if the number of non-cooperative vehicles is very large. For this reason,  security, privacy and safety of VANET users  are the main goals of our design. Hence, VAiPho integrates security mechanisms in order to avoid possible attacks against those characteristics. In the following, several security aspects are discussed from three different points of view: user, information and network.

\subsection{User Security}
\label{sec:UserSecurity}

One of the most important issues in security is anonymity. In VANETs it is undesirable  that the communicating parties reveal their identities and that any attacker could detect a particular phone signal to track somebody. In centralized systems based on data connection, the situation could be worse because the attacker could track all the vehicles through their mobile phone numbers from a single site.  VAiPho mechanism never uses data connection so this centralized attack is impossible.  However, the particular phone signal could be tracked so that in order to provide user anonymity, the proposed system uses variable pseudonyms as identifiers. In particular, each node changes   its pseudonym at random time periods and warns about these changes through  beacons sent only  to those nodes with which  it is authenticated.

Another important security issue consists in ensuring that the device corresponds to a legitimate user on the VAiPho network. Checking this in a fully distributed network, where there is no central infrastructure to control identities, is complex. In this work, the mechanism to verify the authenticity of users  is based on a zero-knowledge proof, which is an interactive method for one user to prove to another user that he/she knows a secret, without revealing anything about it. Specifically, in VAiPho the secret is the public key of a node that both users know, so public keys are  specially selected to fulfill the requirements of the used zero-knowledge proof. In this way, it is necessary that both users at least share a friend  to make authentication possible. Consequently, it is important that users  maintain  their certificate repositories updated. According to the so-called six degrees of separation theory \cite{Papadopouli}, nodes in this type of network have in their local repository at least one node in common with a high probability. VAiPho automatically devaluates the user certificates that show bad behavior,  so that it revokes them  in case of repeated misbehavior. The information about revoked users is also exchanged after the authentication process.

\subsection{Information Security}
\label{sec:InfoAttes}

An attacker could simulate a non-existent traffic congestion in order to convince other users not to choose a route and, in this way, to have the road free of cars. For this reason, VAiPho uses a data aggregation mechanism that helps to avoid this type of  fraud. The first device that  detects an abnormal situation, such as a speed much lower than expected for a long time, sends a traffic congestion warning to its neighboring phones. If these phones detect that their speed is abnormal too, they sign the received information, thus corroborating it. When the promoting phone receives a certain number of signatures, it adds them to an aggregated packet formed with the received information and sends it to all neighbors, who will disseminate the message.

Thus, traffic congestions must be detected by different vehicles, which must sign the traffic event with their private signature in order to aggregate them in a single packet. Therefore, we ensure that not only a vehicle has detected an incident but also several have corroborated it. This mechanism eliminates the possibility of spreading false traffic congestions created by a single attacker, and  avoids possible confusion generated by the system when a vehicle stops at the side of the road due to a flat tire or because the vehicle has  broken down.

The threshold number of required signatures depends on the expansion of the tool, that is to say, the higher the number of vehicles with VAiPho, the greater the number of required signatures. This means that the larger the number of VAiPho devices in the network, the lower the probability of attack. In order to compute the threshold number of required signatures, VAiPho checks the time from its first authentication with other users in the current journey and calculates the average number of users per minute. In the current implementation, this average is lower than one per minute, so the number of required signatures is two. If the average is between one and four, the number of required signatures is four, and if it is higher than four, the number of required signatures is five.

\subsection{Network Security}
\label{sec:NetSecurity}
Nowadays, wireless connectivity in VAiPho is over WLAN but in the future it is expected that will be over Wi-Fi Direct  or by using the IEEE 802.11p standard. The packets are exchanged among vehicles using others as relaying nodes. An attacker could try that communications fail, what could break the VANET into pieces so that it cannot provide services such as packet forwarding. In this sense, these attackers would cause a passive denial-of-service with the goal that the wireless network does not work properly. We must ensure that only those vehicles that belong to the network and help in its operation, benefit from the information relayed in it.  VAiPho prevents users who want to benefit from the VANET without helping in forwarding  information because such a passive attack would degrade  tool functionality and compromise network connectivity.
VAiPho includes a specific mechanism against these attacks, which uses encrypted exchange of data as a method to strengthen cooperation in relaying packets because it prevents passive nodes that do not cooperate in relaying packets from getting benefit from encrypted information.

\section{Simulation and Implementation}
\label{sec:Result}
In this section,     implementation settings and simulation data are briefly described to show that VAiPho is successful in accomplishing the goals of automatically detecting and announcing traffic congestions and helping people to find empty parking spaces. Furthermore, all this is accomplished in a safe way, both regarding received information and user privacy.

The main goal of these proof-of-concept tests was to evaluate whether the deployment of VANETs with mobile phones can fulfill  specific characteristics of VANETs such as  hybrid architecture, high mobility, dynamic topology, scalability problems, and intermittent and unpredictable communications.
The first test consisted of checking that communications between phones with Wi-Fi  using the IEEE 802.11b/g standard are feasible using a simple client-server application. It was tested while  driving in urban environments and highways, at different speeds, and using different number of devices. These tests were successful. After this, we created several simulations of the proposal by using SUMO and NS-2. Finally,  we implemented the complete VAiPho application in smartphones first in Windows Mobile, and then in Android and Symbian.

\subsection{Simulation}

Both the feasibility and the effectiveness of VAiPho approach were shown through the simulation. In the first part of the demonstration with NS-2 \cite{NS2} and SUMO \cite{SUMO} (see Fig. 7), the most relevant  options were:
total number of vehicles: 600 - 15000,
number of vehicles with OBUs: 1\%-100\% ,
simulation time: 100-216000 seconds,
authentication period: 20 seconds, and
distance between relay nodes: 75 meters.

\begin{figure*}
\centering
  \includegraphics[width=0.970\textwidth]{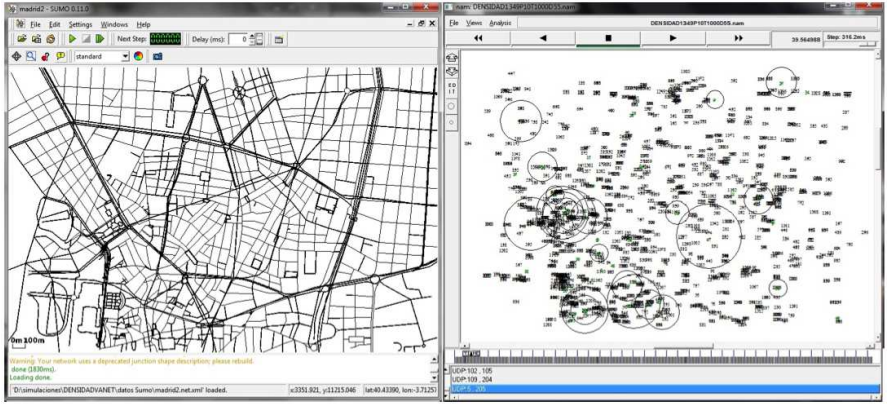}
\caption{VAiPho Simulation with SUMO and NS-2}
\label{fig:SumoNS2}
\end{figure*}

The implemented simulations were analyzed in  three different development layers: vehicle mobility, node energy and P2P communication:

\begin{itemize}
\item The vehicle mobility layer manages the node movement according to the movement pattern, which defines roads, lanes, different speed limits for each lane, traffic congestions, etc. 
\item The node energy layer is used to distinguish between vehicles with and without OBUs because vehicles without OBUs can be  on the road but do not contribute
to the communications. 
\item The P2P communication layer is responsible for the definition of which nodes are in the transmission range of the retransmitting node at any time. 
\end{itemize}

Implemented simulations  let us  know the number of connections related to the percentage of vehicles with OBUs, so from this information  the  minimum  number of vehicles with OBUs necessary  to exchange information can be computed.
On the other hand, the percentage of vehicles with OBUs necessary to prevent degradation in the quality of communications is also computed.

The implemented simulations have been performed for different percentages of nodes with the same topology in order to help to illustrate the vehicular P2P network operation.
Among the  information obtained from the simulations we have the number of packets that have been generated, sent, broadcasted, received, lost, etc. by each node. Also, other interesting pieces of information  are the number of  generated or lost packets  in the whole network, which nodes generate packets and which nodes forward them, etc. Fig. 8 shows the number of packets that were sent and lost during the test. In addition to all this information, another important aspect of the simulation is that it provides a detailed image of what is happening in each moment in the VANET thanks to the use of the NS-2 display. It also shows the traffic model through the SUMO tool while the information is represented using TraceGraph.

\begin{figure*}
\centering
  \includegraphics[width=1\textwidth]{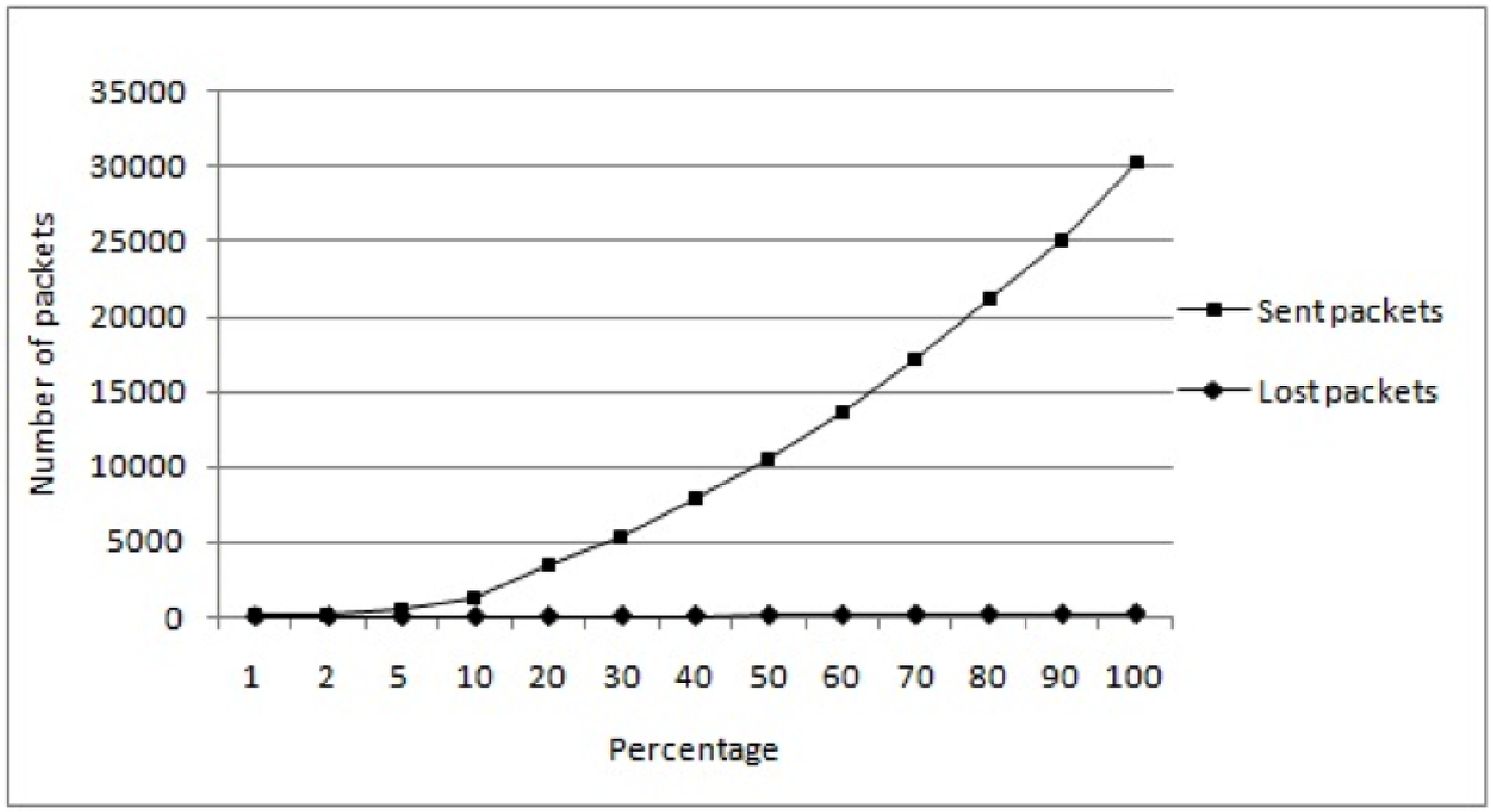}
\caption{Numbers of Packets in VAiPho Simulation}
\label{fig:paquetes}
\end{figure*}

\subsection{Implementation in Smartphones}

For the performance evaluation of the proposal, we first built VAiPho application in different smartphones with Windows Mobile using Microsoft Visual C\# .NET. A video showing how this tool works in real devices, vehicles and roads can be found in the Website \cite{VaiWeb}. 

The tool has shown to be effective with traffic congestion recognition, parking detection and  parking reminder. It also works well regarding  authentication of users and data, protection of user privacy  and exchange of secret information among devices. Also devices running VAiPho display and forward  received information correctly.

In particular tests showed good performance in:
\begin{itemize}
\item Detection and warning of a traffic congestion: A device automatically detects it and sends a warning message,  another device on the same route and direction detects the same traffic congestion and signs the received message, and the signed message is sent to a third device that displays the traffic congestion both on the screen and through voice messages.
\item Parking detection: A vehicle leaves a parking space and sends this information to nearby vehicles so that vehicles receiving this announcement forward the information looking for empty parking spaces show the information on the map.
\item Finding a parked  vehicle: A  vehicle is parked in a parking space and then the user leaves it. Afterwards,  the user presses the parking reminder and the tool shows the walking route  to the vehicle.
\item Geolocation advertising, which is broadcasted in its utility area.
\end{itemize}

In order to test the traffic congestion detection and warning systems, several devices playing  different roles have been used to simulate a real traffic congestion situation. The devices inside vehicles were in charge of detecting, signing and aggregating the event, and receiving those packets and checking the corresponding signatures before warning the driver and relaying this information. 

For the purpose of parking detection, one of the cars was parked and turned  off. After that, we turned the car on and after synchronizing with GPS, the device stored and sent the geographic coordinates to  other devices  in its range of emission. Those vehicles receiving this information that had previously launched the parking detection  application detect and show the empty parking space.

\section{Conclusions}
\label{sec:Conclusion}

This paper proposes a new software tool  for the practical deployment of self-organized VANETs, mainly with the objective of detecting and announcing traffic congestions. 
The proposal, called VAiPho, does not require  any infrastructure neither on the roads nor in vehicles, so it allows a gradual and immediate introduction of VANETs. 
The main goal of this paper is to present VAiPho  algorithms and corresponding software that have been designed to be run on current mobile phones equipped with Wi-Fi connection and GPS. 
One of the main features of the proposed application is that the devices with VAiPho can communicate with each other securely. In particular, its design took into account  integrity and authenticity of   transmitted information  and user privacy. 
Besides,  VAiPho operation is  automatic because it  detects automatically where and when a traffic congestion is  formed. 
Furthermore, the information is provided to the driver through voice warning messages in order to prevent driver distraction. 
VAiPho also provides both the possibility of finding empty parking spaces when the user requests it, and  of locating  a parked vehicle.
 The implementations and simulations of the proposal show that  VAiPho can be effectively deployed in urban situations for specific VANET uses with nowadays mobile phones, while guaranteeing  efficiency and security of communications. Among the works in progress, the implementations for different platforms, like iOS and Windows Phone, and the development of additional functionalities are the most important ones.

\section*{Acknowledgements}
Research supported by the Spanish MINECO and the European FEDER Fund under Projects TIN2011-25452 and IPT-2012-0585-370000, and the FPI scholarship BES-2009-016774.

\end{document}